\documentclass[11pt,a4paper]{article}
\usepackage{amsmath,amssymb,color}
\usepackage{epsfig,graphicx}
\usepackage{scalefnt}
\begin{document}

\title{\bf Precise Charm- and Bottom-Quark Masses: Theoretical and
  Experimental Uncertainties\footnote{Presented at Quarks~2010, 16th
  International Seminar of High Energy Physics, Kolomna, Russia, June 6-12, 2010.}
\footnote{Preprint numbers: SFB/CPP-10-92, TTP10--45, MPP-2010-135}}
\author{K. Chetyrkin$^{a}$,
  J.H. K\"uhn$^a$\footnote{Speaker, e-mail: johann.kuehn@kit.edu},
  A. Maier$^{a}$, P. Maierh\"ofer$^{b}$,\\ P. Marquard$^{a}$,
  M. Steinhauser$^{a}$ and C. Sturm$^{c}$
  \\[.5em]
  $^a$ \small{\em Institut f\"ur Theoretische Teilchenphysik,
    }\\
  {\small\em 
    Karlsruhe~Institute~of~Technology~(KIT), 76128
    Karlsruhe, Germany}
  \\[.3em]
  $^b$ \small{\em Institut f\"ur Theoretische Physik, Universit\"at Z\"urich, 8057
    Z\"urich, Switzerland} 
  \\[.3em]
  $^c$ \small{\em Max-Planck-Institut f{\"u}r Physik,
(Werner-Heisenberg-Institut),
80805 M{\"u}nchen, Germany
} }
\date{}
\maketitle

\begin{abstract}
  Recent theoretical and experimental improvements in the determination
  of charm and bottom quark masses are discussed. A new and improved
  evaluation of the contribution from the gluon condensate $\langle
  \frac{\alpha_s}{\pi} G^2\rangle$ to the charm mass determination and a
  detailed study of potential uncertainties in the continuum cross
  section for $b\bar b$ production is presented, together with a study
  of the parametric uncertainty from the $\alpha_s$-dependence of our
  results.  The final results, $m_c(3\,\text{GeV})=986(13)\,$MeV and
  $m_b(m_b)=4163(16)\,$MeV, represent, together with a closely related
  lattice determination $m_c(3\;{\rm GeV})=986(6)\,$MeV, the presently
  most precise determinations of these two fundamental Standard Model
  parameters. A critical analysis of the theoretical and experimental
  uncertainties is presented.
\end{abstract}

The past years have witnessed significant improvement in the
determination of charm and bottom quark masses as a consequence of
improvements in experimental techniques as well as theoretical
calculations. 
Quark mass determinations can be based on a variety of observables and
theoretical calculations. The one presently most precise follows an idea
advocated by the ITEP group more than thirty years ago
\cite{Shifman:1978bx}. Members of the ITEP group are in the audience and
already for this reason it is appropriate to present the most recent
advances based on this method. Indeed, this approach has gained renewed
interest after significant advances in higher order perturbative
calculations have been achieved.   
In particular the four-loop results (i.e. the coefficients $\bar{C}_n$
discussed  
below) are now available for the Taylor coefficients of the vacuum 
polarization, analytically up to $n=3$ and numerically up to $n=10$.
The method exploits the fact that the vacuum polarization 
function $\Pi(q^2)$ and its derivatives, evaluated at $q^2=0$, can be
considered as 
short distance quantities with an inverse scale characterized by the
distance between the reference point $q^2=0$ and the location of the
threshold $q^2=(3\;{\rm GeV)}^2 $ and $q^2=(10\;{\rm GeV})^2$ for charm and
bottom, respectively. This idea has been taken up in \cite{Kuhn:2001dm} after the
first three-loop evaluation of the moments became available 
\cite{Chetyrkin:1995ii,Chetyrkin:1996cf,Chetyrkin:1997mb} and
has been further improved in \cite{Kuhn:2007vp} using four-loop results
\cite{Chetyrkin:2006xg,Boughezal:2006px} for the
lowest moment. An analysis which included additional theoretical
\cite{Maier:2008he,Maier:2009fz,Kiyo:2009gb,Hoang:2008qy} and experimental results has 
been performed in \cite{Chetyrkin:2009fv}. In the following we
present an improved treatment of the contribution from
the gluon condensate to the moments of the charm correlator and a
critical discussion of $R_b$ in the transition region from the threshold
to the perturbative continuum and the resulting impact on the $m_b$
determination. 

Let us recall some basic notation and definitions. The vacuum
polarization $\Pi_Q(q^2)$ induced by a heavy quark $Q$ with charge
$Q_Q$ (ignoring in this short note the so-called singlet contributions),
is an analytic function with poles and a branch cut at and above
$q^2=M_{J/\psi}^2$ for charm (or $M_\Upsilon^2$ for bottom). Its
Taylor coefficients $\bar C_n$, defined through 
\begin{equation}
  \label{eq:Pi}
  \Pi_Q(q^2)\equiv Q_Q^2\frac{3}{16\pi^2} \sum_{n\geq0}\bar{C}_nz^n
  \,,
\end{equation}
can be evaluated in perturbative QCD (pQCD), presently up to order $\alpha_s^3$.
Here $z\equiv q^2/(4m_Q^2)$, where $m_Q=m_Q(\mu)$ is the running
$\overline{\text{MS}}$ mass at scale $\mu$. Using a once-subtracted dispersion
relation
\begin{equation}
  \label{eq:disp_rel}
  \Pi_Q(q^2)=\frac{q^2}{12\pi^2}\int {\rm d}s \frac{R_Q(s)}{s(s-q^2)}
  \,,
\end{equation}
(with $R_Q$ denoting the familiar $R$-ratio for the production of heavy quarks 
with flavor $Q$), the Taylor coefficients can be expressed through moments of 
$R_Q$. Equating perturbatively calculated and experimentally measured moments,
\begin{equation}
  \label{eq:M_exp}
  {\cal M}_n^{\text{exp}}=\int \frac{{\rm d}s}{s^{n+1}}R_Q(s)
  \,,
\end{equation}
leads to an ($n$-dependent) determination of the quark mass
\begin{equation}
\label{eq:m_Q}
m_Q= 
\frac{1}{2}\left(
\frac{9Q_Q^2}{4}\frac{\bar{C}_n}{{\cal M}_n^{\text{exp}}}
\right)^{\frac{1}{2n}}\,.
\end{equation}

Significant progress has been made in the perturbative evaluation of
the moments since the first analysis of the ITEP group. The 
${\cal O}(\alpha_s^2)$ contribution (three loops) has been evaluated more
than 13 years ago \cite{Chetyrkin:1995ii,Chetyrkin:1996cf,Chetyrkin:1997mb}, 
as far as the terms up to $n=8$ are
concerned, recently even up to $n=30$ \cite{Boughezal:2006uu,Maier:2007yn}.  
About ten years later the
lowest two moments ($n=0,1$) of the vector correlator were evaluated
in ${\cal O}(\alpha_s^3)$, i. e. in four-loop approximation
\cite{Chetyrkin:2006xg,Boughezal:2006px}. 
The corresponding two lowest moments for the pseudoscalar
correlator were obtained in \cite{Sturm:2008eb} in order to derive the charmed
quark mass from lattice simulations \cite{Allison:2008xk}.  
In \cite{Maier:2008he,Maier:2009fz} the second and
third moments were evaluated for vector, axial and pseudoscalar
correlators. Combining, finally, these results with information about
the threshold and high-energy behavior in the form of a Pad\'e
approximation, the full $q^2$-dependence of all four correlators was
reconstructed and the next moments, from four up to ten, were obtained
with adequate accuracy~\cite{Kiyo:2009gb} (see also Ref.~\cite{Hoang:2008qy}).

Most of the experimental input had already been compiled and exploited
in \cite{Kuhn:2007vp} both for the charm and the bottom case. In
\cite{Kuhn:2007vp,Chetyrkin:2009fv} an estimate for the gluon condensate
was included which gives a tiny contribution for the charm case. This
estimate for the moments was based on the results from~\cite{Novikov:1977dq}
plus  next-to-leading order (NLO) terms from~\cite{Broadhurst:1994qj}
\begin{equation}
  \label{eq:1}
  \delta {\cal M}^{\text{np}}_n = \frac{12 \pi^2 Q_c^2}{(4
    m_{c,\text{pole}}^2 )^{n+2}} \left\langle \frac{\alpha_s}{\pi} G^2
  \right\rangle a_n (1+\frac{\alpha_s}{\pi} b_n ) ,
\end{equation}
with 
\begin{eqnarray}
  &&a_n = - \frac{2n+2}{15}
  \frac{\Gamma(4+n)\Gamma\left(\frac{7}{2}\right)}
  {\Gamma(4)\Gamma\left(\frac{7}{2}+n\right)}\,,
  \nonumber\\
  &&\quad b_1 = \frac{135779}{12960}\,, \quad b_2 = \frac{1969}{168}\,,
  \quad b_3 = \frac{546421}{42525}\,, \quad b_4 = \frac{661687433}{47628000}\,.
\end{eqnarray}
Using the ${\cal O}(\alpha_s)$ pole-$\overline{\text{MS}}$-mass
conversion , $m_Q(\mu)/m_{Q,\text{pole}} =
1-\alpha_s/\pi\times[4/3+ \ln(\mu^2/m_Q^2)]$, Eq.~(\ref{eq:1})
was expressed in 
terms of the $\overline{\text{MS}}$ mass at the scale $\mu = 3 \,
\text{GeV}$ with new coefficients 
$\bar b_n = b_n - (2n + 4)(4/3+ \ln(\mu^2/m_Q^2))$. 
However, as a consequence of the high inverse
power of the mass appearing in Eq.~(\ref{eq:1}) this conversion becomes
unstable.\footnote{We would like to thank S.~Bodenstein for drawing our attention
  to this point.} (Although the effect obtained in
\cite{Kuhn:2007vp,Chetyrkin:2009fv} was small, nevertheless.) For this
reason we prefer to use directly the on-shell formulation, Eq.~(\ref{eq:1})
with the parameters $m_{c,\text{pole}} = 1.5$~GeV,
$\alpha_s^{(4)}(3~\mbox{GeV}) = 0.258$ 
(obtained from $\alpha_s^{(5)}(M_Z)=0.1189$) and 
$\langle \frac{\alpha_s}{\pi} G^2 \rangle = 0.006\pm 0.012\,
\text{GeV}^4$~\cite{Ioffe:2005ym}. The contributions to the moments are
displayed in Tab.~\ref{tab:contglu} both for the LO and the NLO predictions,
together with 
the contribution from the narrow resonances, threshold and continuum,
which are directly copied from \cite{Kuhn:2007vp}. The results for the
charm mass, $m_c(3\,\text{GeV})$, are shown in Tab.~\ref{tab:charmTab}
again for the two choices of the condensate contribution.

  \begin{table}[t]
    \centering
\begin{tabular}{l|lll|l||ll}
\hline
$n$ & ${\cal M}_n^{\rm res}$
& ${\cal M}_n^{\rm thresh}$
& ${\cal M}_n^{\rm cont}$
& ${\cal M}_n^{\rm exp}$
& ${\cal M}_n^{\rm np}$(LO)& ${\cal M}_n^{\rm np}$(NLO)
\\
& $\times 10^{(n-1)}$
& $\times 10^{(n-1)}$
& $\times 10^{(n-1)}$
& $\times 10^{(n-1)}$
& $\times 10^{(n-1)}$
\\
\hline
$1$&$  0.1201(25)$ &$  0.0318(15)$ &$  0.0646(11)$ &$  0.2166(31)$
&$-0.0001(3)$ &$-0.0002(5)$ \\
$2$&$  0.1176(25)$ &$  0.0178(8)$ &$  0.0144(3)$ &$  0.1497(27)$
&$-0.0002(5)$ &$-0.0005(10)$ \\
$3$&$  0.1169(26)$ &$  0.0101(5)$ &$  0.0042(1)$ &$  0.1312(27)$
&$-0.0004(8)$ &$-0.0008(16)$ \\
$4$&$  0.1177(27)$ &$  0.0058(3)$ &$  0.0014(0)$ &$  0.1249(27)$
&$-0.0006(12)$ &$-0.0013(25)$ \\
\hline
\end{tabular}
\caption{Experimental moments
  in $(\mbox{GeV})^{-2n}$ as defined in
  Eq.~(\ref{eq:M_exp}), separated according to the contributions from
  the narrow resonances,
  the charm threshold region and the continuum region
  above $\sqrt{s}=4.8$~GeV. In the last two columns the contribution from the
  gluon condensate in LO and NLO are shown.} 
    \label{tab:contglu}
  \end{table}

  \begin{table}[t]
    \centering\scalefont{1}{
\begin{tabular}{l|l|l|lllll|ll}
\hline
$n$ & $m_c(3~\mbox{GeV}) $ & 
$m_c(3~\mbox{GeV}) $ &
exp & $\alpha_s$ & $\mu$ & ${\rm np}_{\rm LO}$& ${\rm np}_{\rm NLO}$ &
total   & total 
\\
 & $[{\rm np}_{\rm LO}]$ & $[{\rm np}_{\rm NLO}]$ &
 &  & & & &
 $[{\rm np}_{\rm LO}]$  & $[{\rm np}_{\rm NLO}]$
\\
\hline
1&  0.986& 0.986& 0.009&  0.009&  0.002&  0.001& 0.001& 0.013&0.013\\
2&  0.976& 0.975& 0.006&  0.014&  0.005&  0.001&  0.002& 0.016& 0.016\\
3&  0.976& 0.975& 0.005&  0.015&  0.007&  0.001&  0.003&0.017& 0.017\\
4&  1.000& 0.999& 0.003&  0.009&  0.031&  0.001&   0.003&0.032& 0.032\\
\hline
\end{tabular}}
    \caption{Results for $m_c(3\, \text{GeV} )$ in $\text{GeV}$
      including the LO or NLO order gluon condensate contribution. The errors
      are from experiment, $\alpha_s$, variation of $\mu$ and the
      different options for the gluon condensate.}
    \label{tab:charmTab}
  \end{table}

As is evident from Tabs.~\ref{tab:contglu} and~\ref{tab:charmTab}, the
effect of the gluon condensate remains small, in particular for the
lowest three moments. The final result 
\begin{eqnarray}
m_c (3 \, \text{GeV}) = 986(13)\, \text{MeV}
\,,
\end{eqnarray}
remains unchanged when compared to
\cite{Kuhn:2007vp,Chetyrkin:2009fv}. The consistency of this result with
the ones for $n=2,3$ and $4$ can be considered 
as additional
confirmation. Transforming this to the scale-invariant mass 
$m_c(m_c)$~\cite{Chetyrkin:2000yt}, including the
four-loop coefficients  of the renormalization group functions one
finds~\cite{Chetyrkin:2009fv}
$m_c(m_c)=1279(13)~\text{MeV}$. Let us recall at this point that a recent
study~\cite{McNeile:2010ji}, 
combining a lattice simulation for 
the pseudoscalar correlator
with the perturbative  three- and four-loop result
\cite{Chetyrkin:1997mb,Sturm:2008eb,Maier:2009fz} has led to
$m_c(3~\text{GeV})= 986(6)$~MeV in remarkable
agreement with~\cite{Kuhn:2007vp,Chetyrkin:2009fv}.  
Fair agreement is also found with a recent analysis based on finite energy
sum rules using results up to four-loop order in perturbation
theory~\cite{Bodenstein:2010qx} which leads to the result 
$m_c(3~\text{GeV})=1008(26)$~MeV.
A combination of fixed-order moments and effective-theory
calculations has been considered in Ref.~\cite{Signer:2008da}
with the result
$m_c(m_c)=1250(40)$~MeV which corresponds to
$m_c(3~\text{GeV})=966(40)$~MeV.

\begin{figure}[t]
  \centering
  \includegraphics[width=\textwidth]{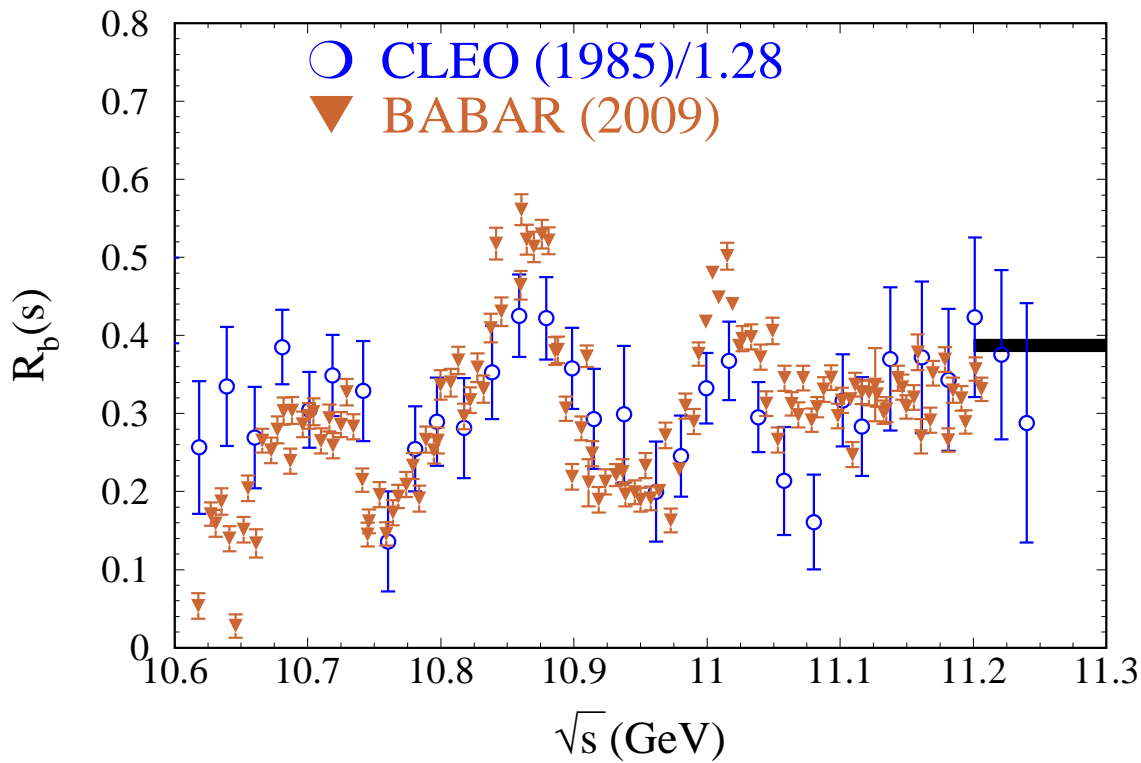}
  \caption{Comparison of rescaled CLEO data for $R_b$ with BABAR data.
    \cite{Chetyrkin:2009fv,:2008hx}. The black bar on the right corresponds to
    the theory prediction~\cite{Harlander:2002ur}.}
  \label{fig:R_b}
\end{figure}

\begin{figure}[t]
  \centering
  \includegraphics[width=\textwidth]{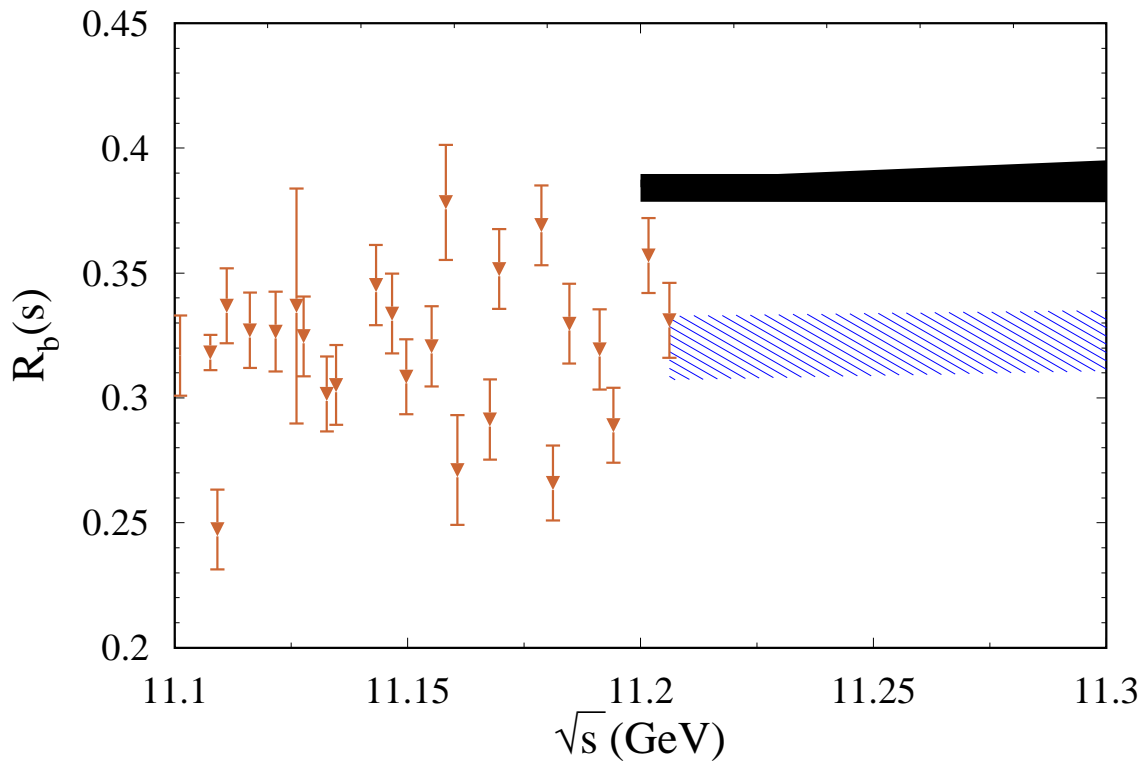}
  \caption{The region around $\sqrt{s}=11.2$~GeV of Fig.~\ref{fig:R_b} is
    magnified. Only the BABAR data is shown. The shaded band corresponds to a
    linear interpolation between $R(11.2062~\mbox{GeV})$ and
    $R(13~\mbox{GeV})$ as described in the text, see option~A.
    The black bar on the right corresponds to
    the theory prediction~\cite{Harlander:2002ur}.}
  \label{fig:bottomInter}
\end{figure}

Until about two years ago the only measurement of the cross section above but
still close to the $B$-meson threshold, i.e. in the region between
$10.6$ and $11.2 \, \text{GeV}$, had been performed by the CLEO collaboration
in the mid-eighties \cite{Besson:1984bd}. Its large systematic
uncertainty was responsible for a sizable fraction of the final error on
$m_b$ in the analysis of \cite{Kuhn:2007vp}.  This measurement had been
superseded by a measurement of BABAR \cite{:2008hx} with a
systematic error around 3\%. In \cite{Chetyrkin:2009fv} the
radiative corrections were unfolded and used to obtain a significantly
improved determination of the moments. These BABAR data are shown in
Fig.~\ref{fig:R_b} together with the theory prediction based on pQCD in
${\cal O}(\alpha_s^2)$. 
Observing that $R_b$ flattens off above $11.1\,\text{GeV}$ one should
expect agreement between pQCD and experiment. The result for the region
above $11.1\,\text{GeV}$ is shown in Fig.~\ref{fig:bottomInter}, again
with the theory prediction.

Taking the average of the data points above 11.1~GeV 
one finds $\overline{R}_b = 0.32$
with negligible statistical and uncorrelated systematical errors. 
The correlated systematical error is quoted to be 3.5\%. 
In \cite{Kuhn:2007vp} and \cite{Chetyrkin:2009fv} data and pQCD were
taken at face value for $\sqrt{s}$ below and above 11.2~GeV,
respectively (with linear interpolation between the last data point
$R_b(11.2062 \,\text{GeV}) = 0.331$ and the pQCD prediction
$R_b^{\text{pQCD}}(11.24\,\text{GeV})=0.387$). This leads to the moments
and the quark masses as shown in Tabs.~\ref{tab:mbMomOld} and~\ref{tab::mb}, 
respectively. 

\begin{table}[t]
  \centering
  \begin{tabular}{c|lll|l}
\hline
$n$&${\cal M}_n^{\text{res,(1S-4S)}}$&${\cal M}_n^{\text{thresh}}$&${\cal M}_n^{\text{cont}}$&${\cal M}_n^{\text{exp}}$
\\
& $\times 10^{(2n+1)}$
& $\times 10^{(2n+1)}$
& $\times 10^{(2n+1)}$
& $\times 10^{(2n+1)}$\\
\hline
    $1$&$   1.394(23)$ &$   0.287(12)$ &$   2.911(18)$ &$   4.592(31)$ \\
$2$&$   1.459(23)$ &$   0.240(10)$ &$   1.173(11)$ &$   2.872(28)$ \\
$3$&$   1.538(24)$ &$   0.200(8)$ &$   0.624(7)$ &$   2.362(26)$ \\
$4$&$   1.630(25)$ &$   0.168(7)$ &$   0.372(5)$ &$   2.170(26)$ \\\hline
  \end{tabular}
  \caption{Moments for the bottom quark system in $(\text{GeV})^{-2n}$}
  \label{tab:mbMomOld}
\end{table}

\begin{table}[t]
\begin{center}
{

\begin{tabular}{l|l|rrr|l|l}
\hline
$n$ & $m_b(10~\text{GeV})$ & 
exp & $\alpha_s$ & $\mu$ &
total &$m_b(m_b)$
\\
\hline
        1&  3.597&  0.014&  0.007&   0.002&  0.016&  4.151 \\
        2&  3.610&  0.010&  0.012&  0.003&  0.016&  4.163 \\
        3&  3.619&  0.008&  0.014&   0.006&  0.018&  4.172 \\
        4&  3.631&  0.006&  0.015&  0.020&  0.026&  4.183 \\
\hline
\end{tabular}
}
\caption{\label{tab::mb}Results for $m_b(10~\text{GeV})$ and $m_b(m_b)$ 
in GeV
  obtained from Eq.~\eqref{eq:m_Q}.
  The errors are from experiment, $\alpha_s$ and the variation of $\mu$.
}
\end{center}
\end{table}

This procedure is based on the assumption that pQCD is valid in
  the region above \mbox{$\sim11.2$~GeV}. where the relative momentum of
  $b$ and $\bar{b}$ quarks has reached about 5~GeV. Indeed,
  considering the behaviour of $R_b$ as displayed in
  Fig.~\ref{fig:R_b} it is quite plausible that $R_b$ quickly reaches
  the level anticipated by pQCD. However, considering the $20$\%
  deviation between data and pQCD around 11.2~GeV one may consider
  the possibility that either pQCD is valid only at significantly
  higher energies (option~A), or that the systematic error of the
  BABAR data is significantly underestimated, requiring a shift of the
  data by a sizeable amount (option~B). These options should be
  considered to be ``worst case'' scenarios. Nevertheless we shall
  demonstrate that the resulting shifts are only slightly larger than
  the error quoted in Ref.~\cite{Chetyrkin:2009fv}. Let us now explore
  these two options in detail:
\begin{description}
\item[Option A:] pQCD is only valid at higher energies, say above
  $13~\text{GeV}$ with linear interpolation between
  {$R_b(11.2~\text{GeV})=0.32$ and $R^{\rm pQCD}_b(13~\text{GeV})=0.377$. 
  The results for the moments and $m_b(10~\text{GeV})$ are shown in 
  Tabs.~\ref{tab:m_b_mom_v1} and~\ref{tab:m_b_v1}, respectively,
  assuming a 4\% uncertainty at $\sqrt{s}=11.2062$~GeV
  and no uncertainty for $R(\sqrt{s}=13~\mbox{GeV})$.
  A remarkable stability is observed for the bottom quark mass as can
  be seen from Tab.~\ref{tab:m_b_v1}. For $n=2$ one obtains 
  $m_b(10~\mbox{GeV})=3.630$~GeV.
  }

\item[Option B:] Perturbative QCD is valid at 11.2~GeV, and the systematic
  error of~\cite{:2008hx} is assumed to be underestimated. Therefore the
  {data in the threshold region are rescaled by a factor 
  $R^{\rm pQCD}_b/\overline{R}_b = 0.387/0.32 \approx 1.21$
  corresponding to a shift of about 7~$\sigma$.
  The results for the moments are shown in Tab.~\ref{tab:m_b_mom_v2}, the
  corresponding predictions for $m_b(10\, \text{GeV})$ in
  Tab.~\ref{tab:m_b_v2}, with $m_b(10~\mbox{GeV})=3.592$~GeV for $n=2$.
  As expected, the trend of increasing $m_b$ with increasing $n$
  already present in Tab.~\ref{tab::mb} is
  emphasized even more.
}
\end{description}

\begin{table}[t]
   \centering
\begin{tabular}{l|lll|l}
\hline
$n$
& ${\cal M}_n^{\rm res,(1S-4S)}$
& ${\cal M}_n^{\rm thresh}$
& ${\cal M}_n^{\rm cont}$
& ${\cal M}_n^{\rm exp}$
\\
& $\times 10^{(2n+1)}$
& $\times 10^{(2n+1)}$
& $\times 10^{(2n+1)}$
& $\times 10^{(2n+1)}$
\\
\hline
$1$&$   1.394(23)$ &$   0.270(11)$ &$   2.854(17)$ &$   4.518(31)$\\
$2$&$   1.459(23)$ &$   0.226(9)$ &$   1.133(11)$ &$   2.819(27)$ \\
$3$&$   1.538(24)$ &$   0.190(8)$ &$   0.596(8)$ &$   2.324(26)$ \\
$4$&$   1.630(25)$ &$   0.159(6)$ &$   0.353(5)$ &$   2.142(26)$ \\
\hline
\end{tabular}
\caption{Moments for the bottom quark system in $(\text{GeV})^{-2n}$ obtained
  using option~A. The contribution from the linear interpolation
  is contained in ${\cal M}_n^{\rm cont}$.} 
   \label{tab:m_b_mom_v1}
 \end{table}

 \begin{table}[t]
   \centering
   \begin{tabular}{l|c|cccc|c}
\hline
$n$& $m_b(10~\mbox{GeV})$& exp& $\alpha_s$ & $\mu$  &
total& $m_b(m_b)$ \\\hline
 1&  3.631&  0.014&  0.007&  0.002&  0.016&  4.183 \\
 2&  3.630&  0.010&  0.012&  0.003&  0.016&  4.182 \\
 3&  3.631&  0.008&  0.014&  0.006&  0.018&  4.183 \\
 4&  3.637&  0.007&  0.015&  0.020&  0.026&  4.189 \\
\hline
   \end{tabular}
   \caption{Bottom quark mass in~$\text{GeV}$ obtained using the option~A.   
     The uncertainties are from experiment, $\alpha_s$ and the variation of $\mu$.}
   \label{tab:m_b_v1}
 \end{table}


\begin{table}[t]
   \centering
\begin{tabular}{l|lll|l}
\hline
$n$
& ${\cal M}_n^{\rm res,(1S-4S)}$
& ${\cal M}_n^{\rm thresh}$
& ${\cal M}_n^{\rm cont}$
& ${\cal M}_n^{\rm exp}$
\\
& $\times 10^{(2n+1)}$
& $\times 10^{(2n+1)}$
& $\times 10^{(2n+1)}$
& $\times 10^{(2n+1)}$
\\
\hline
 $1$&$   1.394(23)$ &$   0.347(14)$ &$   2.911(18)$ &$   4.651(32)$\\
 $2$&$   1.459(23)$ &$   0.290(12)$ &$   1.173(11)$ &$   2.921(28)$ \\
 $3$&$   1.538(24)$ &$   0.242(10)$ &$   0.624(7)$ &$   2.404(27)$ \\
 $4$&$   1.630(25)$ &$   0.203(8)$ &$   0.372(5)$ &$   2.205(27)$ \\
\hline
\end{tabular}
\caption{Moments for the bottom quark system in $(\text{GeV})^{-2n}$ obtained
  using option~B.} 
   \label{tab:m_b_mom_v2}
 \end{table}


 \begin{table}[t]
   \centering
   \begin{tabular}{l|c|cccc|c}
\hline
$n$& $m_b(10~\mbox{GeV})$& exp& $\alpha_s$ & $\mu$  &
total& $m_b(m_b)$ \\\hline
 1&  3.570&  0.015&  0.008&  0.002&  0.017&  4.124 \\
 2&  3.592&  0.010&  0.012&  0.003&  0.016&  4.146 \\
 3&  3.607&  0.008&  0.014&  0.006&  0.018&  4.160 \\
 4&  3.622&  0.006&  0.015&  0.020&  0.026&  4.175 \\
\hline
   \end{tabular}
   \caption{Bottom quark mass in~$\text{GeV}$ obtained using the option~B.   
     The uncertainties are from experiment, $\alpha_s$ and the variation of $\mu$.}
   \label{tab:m_b_v2}
 \end{table}


Similar to the charm case, the result for the bottom mass based on the
lower moments is more stable than the one from moments $n=4$ and above.
In order to suppress the theoretically evaluated
input above 11.2 GeV (which corresponds to roughly 60\% for the lowest,
40\% for the second and 26\% for the third moment), the result from the
second moment has been adopted as our final result,
\begin{equation}
  m_b(10~\text{GeV}) =3610(16)~\text{MeV},
\label{eq:m_b}
\end{equation}
corresponding to $m_b(m_b)=4163(16){\rm MeV}$. 
{Note that options A and
B are considered as ``worst case'' scenarios, nevertheless, the shift
(for $n=2$) is $+20$~MeV and $-18$~MeV, respectively,  
and thus only slightly higher than the uncertainty of 16~MeV.
For this reason we stick to the original result of Eq.~(\ref{eq:m_b}).
}
The explicit $\alpha_s$
dependence of $m_c$ and $m_b$ can be found in
\cite{Chetyrkin:2009fv}. When considering the ratio of charm and bottom
quark masses, part of the  $\alpha_s$ and of the $\mu$ dependence cancels
\begin{equation}
  \label{eq:m_ratio}
  \frac{m_c(3~\text{GeV})}{m_b(10~\text{GeV})}=
  0.2732 -\frac{\alpha_s(M_Z)-0.1189}{0.002}\cdot 0.0014 \pm 0.0028 
  \,.
\end{equation}
This combination might be a useful input in ongoing analyses of bottom decays.

\begin{figure}[t]
\begin{center}
\begin{tabular}{c}
\hspace{-1.cm}
\leavevmode
\epsfxsize=1.0\textwidth
\epsffile{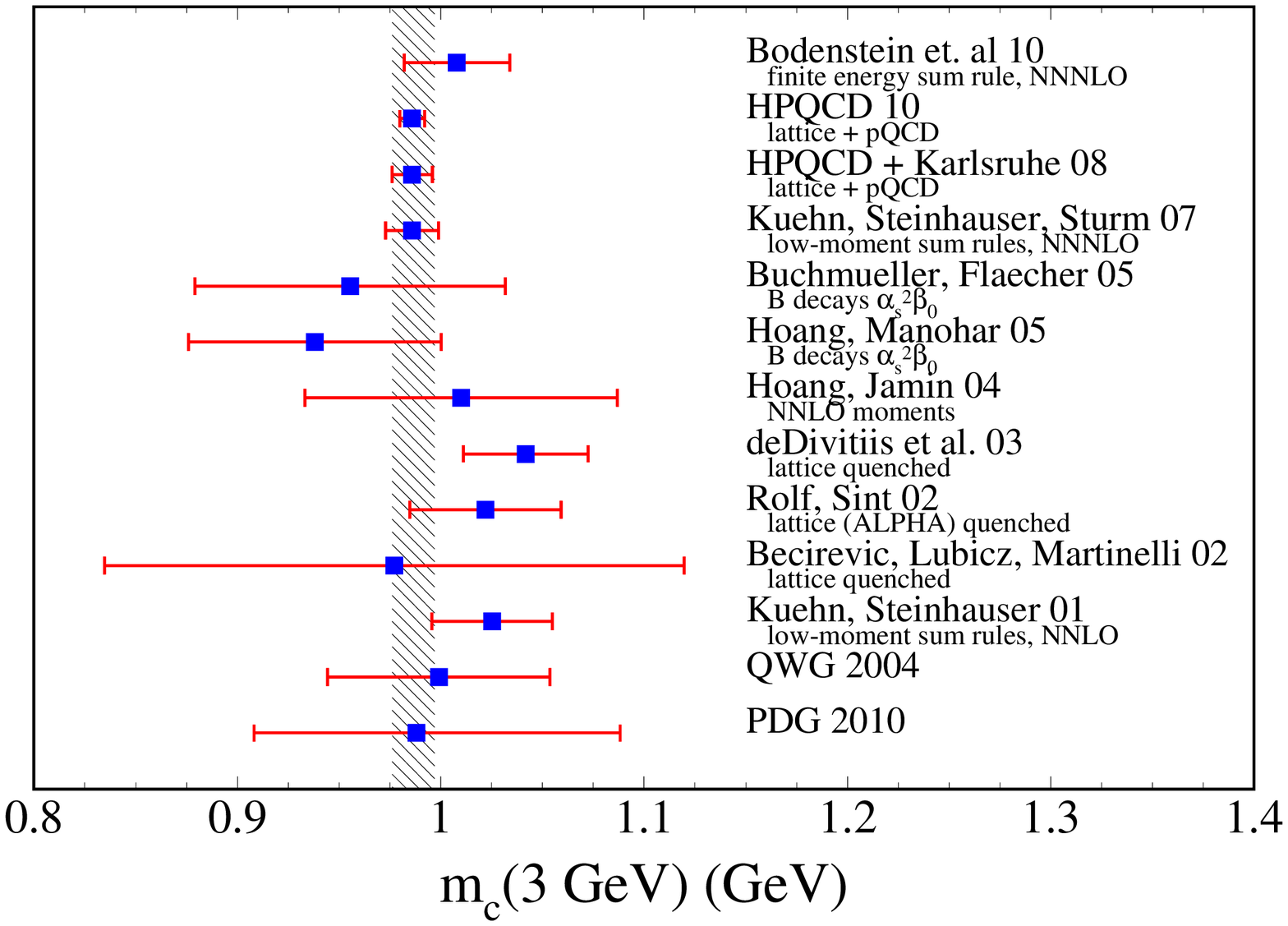}
\\
\hspace{-1.cm}
\leavevmode
\epsfxsize=1.0\textwidth
\epsffile{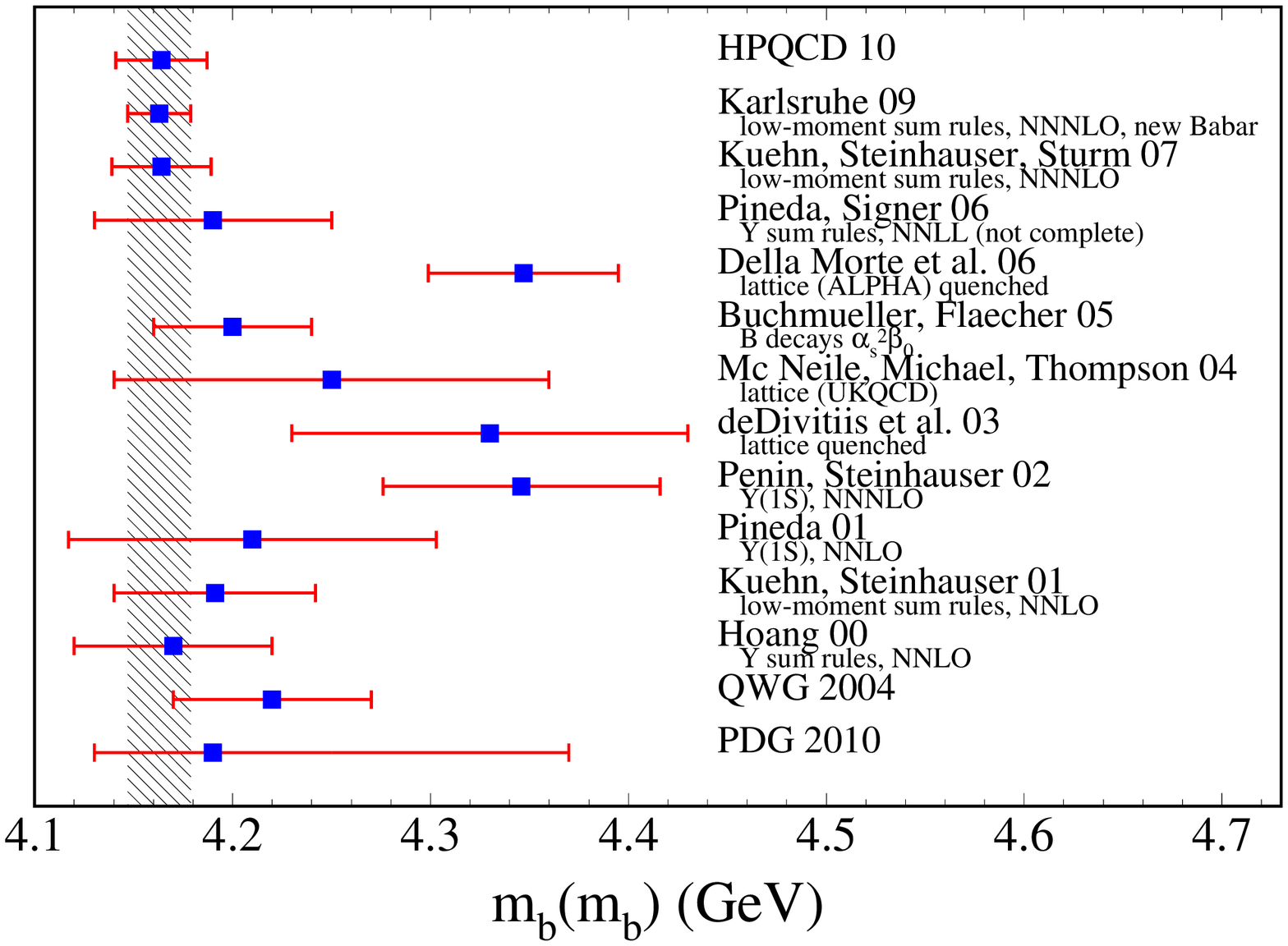}
\end{tabular}
\end{center}
\vspace*{-2em}
\caption{
Comparison of recent determinations of $m_c(3\;{\rm GeV})$ and $m_b(m_b)$.
Shaded band:~\cite{Chetyrkin:2009fv}.
}
\label{fig::mc_compare}
\end{figure}

In Fig.~\ref{fig::mc_compare} the results of this analysis are 
compared to others based on completely different methods. The $m_c$ value is well 
within the range suggested by other determinations. In case of $m_b$ our result 
is somewhat towards the low side, although still consistent with most other 
results.

The results presented in~\cite{Chetyrkin:2009fv} constitute the most 
precise values for the
charm- and bottom-quark masses available to date.\footnote{For the
  charm quark a slightly more precise result has been obtained in
  Ref.~\cite{McNeile:2010ji}.}
Nevertheless it is
tempting to point to the dominant errors and thus identify potential
improvements.  In the case of the charmed quark the error is dominated
by the parametric uncertainty in the strong coupling
$\alpha_s(M_Z)=0.1189\pm0.002$. Experimental and theoretical errors are
comparable, the former being dominated by the electronic width of the
narrow resonances. In principle this error could be further reduced by
the high luminosity measurements at BESS III. A further reduction of the
(already tiny) theory error, e.g.~through a five-loop calculation
looks difficult. Further confidence in our result can be obtained from
the comparison with the aforementioned lattice evaluation.

Improvements in the bottom quark mass determination could originate from
the experimental input, e.g.\ through an improved determination of the
electronic widths of the narrow $\Upsilon$ resonances or through a
second, independent measurement of the $R$-ratio in the region from the
$\Upsilon(4S)$ up to 11.2 GeV. The slight mismatch between the theory
prediction above 11.2 GeV and the data in the region below with their
systematic error of about 3\% has been discussed above. An independent
measurement in the continuum region, e.g. by the BELLE collaboration,
would be extremely important.

In this connection it may be useful to collect the most important pieces of
evidence supporting this remarkably small error. 
Part of the discussion is applicable to both
charm and bottom, part is specific to only one of them. 
In particular for charm, but to some extent also for bottom, the $\mu$
dependence
of the result increases for the higher moments, starting with $n=4$, and 
dominates the total error. We will therefore concentrate on the moments 
$n=1$, 2 and 3 which were used for the mass determination, 
results for $n=4$ will only be mentioned for illustration. 

\begin{figure}[ht]
\begin{center}
  \begin{tabular}{c}
    \leavevmode
    \epsfxsize=1.0\textwidth
    \epsffile{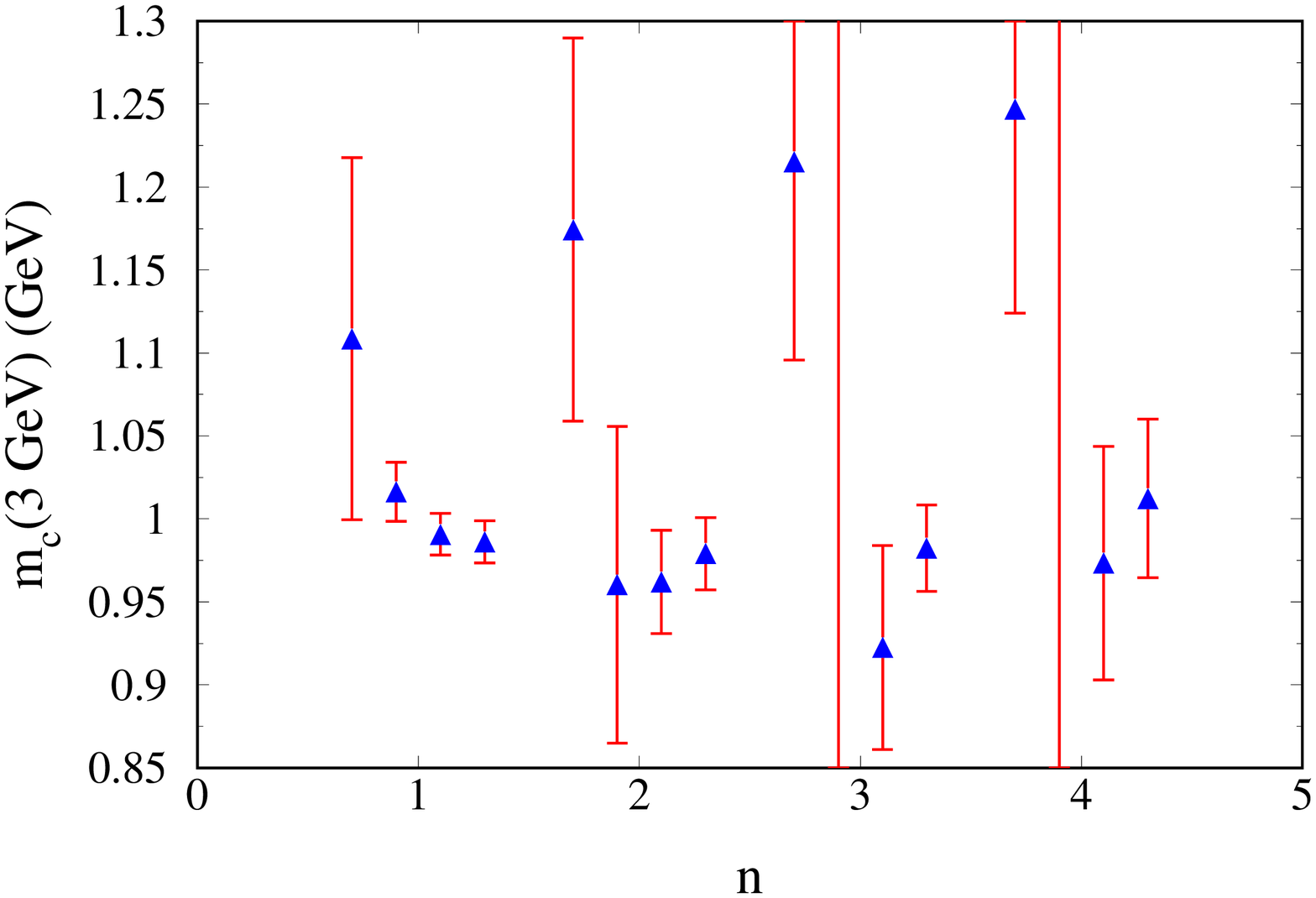}
  \end{tabular}
\end{center}
\vspace*{-2em}
\caption{
$m_c(3~\mbox{GeV})$ for $n=1,2,3$ and $4$.
For each value of $n$ the results from left to right correspond
the inclusion of terms of order $\alpha_s^0$, $\alpha_s^1$,
$\alpha_s^2$ and $\alpha_s^3$.
}
\label{fig::mom}
\end{figure}

Let us start with charm. Right at the beginning it should be
emphasized that the primary quantity to be determined is the 
running mass at the scale of 3~GeV, 
the scale characteristic for the production threshold and thus for
the process. Furthermore, at this scale the strong coupling 
$\alpha_s^{(4)}(3\;{\rm GeV}) =  0.258$ is already
sufficiently small such that the higher order terms in the perturbative series
decrease rapidly. Last not least, for many other observables of interest, like
$B$-meson decays into charm, or processes involving virtual charm quarks like
$B\to X_s\gamma$ or $K\to \pi \nu\bar \nu$, the characteristic scale is of order
3~GeV or higher. Artificially running the mass first down to 
${\cal O} (1\;{\rm GeV})$ and then back to a higher scale thus leads to an
unnecessary inflation of the error.

The quark mass determination is affected by the theory uncertainty, 
resulting in particular from our ignorance of yet uncalculated higher orders,
and by the error in the evaluation of the experimental moments. 
The former has been estimated \cite{Kuhn:2007vp}
by evaluating $m_c(\mu)$ at different renormalization scales
between 2 and 4~GeV (changing, of course, the coefficients $\bar{C}_n$
appropriately) and subsequently evolving $m_c(\mu)$ to $m_c(3\;{\rm GeV})$. The
error estimates based on these considerations are listed in Tab.~\ref{tab:charmTab}.

The stability of the result upon inclusion of higher orders is also evident from 
Fig.~\ref{fig::mom} where the results from different values of $n$ 
are displayed separately in order $\alpha_s^i$ with $i=0$, 1, 2 and 3.
This argument can be made more quantitatively by rewriting eq.~(\ref{eq:m_Q}) 
in the form\footnote{QED corrections are contained in $C_n^{\rm Born}$.}
\begin{eqnarray}
m_c &=& 
\frac{1}{2}
\bigg( \frac{9 Q_c^2}{4}\frac{\bar{C}_n^{\rm Born}}{{\cal M}_n^{\rm exp}}\bigg)^\frac{1}{2n} 
(1 + r^{(1)}_n \alpha_s + r^{(2)}_n\alpha_s^2 + r^{(3)}_n \alpha_s^3) 
\nonumber\\
&\propto&
1 - 
\left(\begin{array}{c}0.328\\0.524\\0.618\\0.662\end{array}\right)\alpha_s - 
\left(\begin{array}{c}0.306\\0.409\\0.510\\0.575\end{array}\right)\alpha_s^2 -
\left(\begin{array}{c}0.262\\0.230\\0.299\\0.396\end{array}\right)\alpha_s^3,
\label{mc} 
\end{eqnarray}
where the entries correspond to the moments with $n=1$, 2, 3 and 4. 
Note, that the coefficients are decreasing with increasing order of
$\alpha_s$. Estimating
the relative error through $r_n^{\rm max}(\alpha_s(3\;{\rm GeV}))^4$ leads to
1.4 /  2.3 / 2.7 / 2.9  per mille and thus to an estimate clearly
smaller than the one based on the $\mu$ dependence. 

\begin{figure}[t]
\begin{center}
  \begin{tabular}{c}
    \leavevmode
    \epsfxsize=1.0\textwidth
    \epsffile{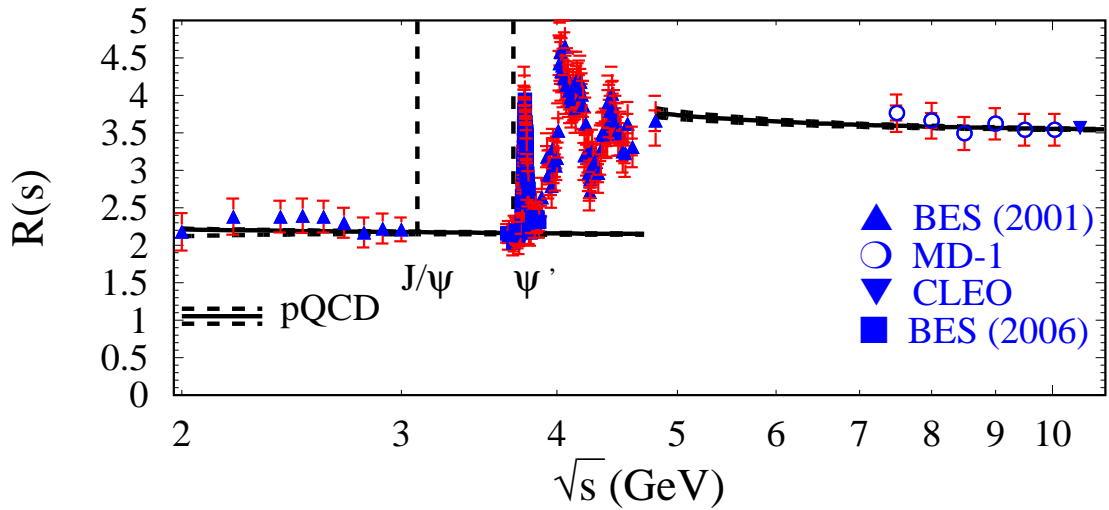}
    \\[-2em]
  \end{tabular}
\end{center}
\caption{$R(s)$ for different energy intervals around the charm threshold 
region. The solid line corresponds to the theoretical prediction.
}
\label{fig::R}
\end{figure}

The consistency between the results for different values of $n$ is another
piece of evidence (Fig.~\ref{fig::mom} and Tab.~\ref{tab:charmTab}). 
For the lowest three moments the variation between the
maximal and the minimal value amounts to 10~MeV only. This, in addition,
points to the selfconsistency of our data set. Let us illustrate this
aspect by a critical discussion of the ``continuum contribution'',  i.e. the
region above 4.8 GeV, where data points are available at widely separated
points only. Instead of experimental data the theory prediction for $R(s)$
has been employed for the evaluation of the contribution to the moments.
If the true contribution from this region would be shifted down by, say,
10\%, this would move $m_c$, as derived from
$n=1$, up by about 20~MeV.  However, this same shift would lead to a
small increase by 3~MeV for  $n=2$ and leave the results 
for higher $n$  practically unchanged.
Furthermore, theory predictions and measurements in the region from
4.8~GeV up to the bottom-meson threshold, wherever available,
are in excellent agreement, as shown in Fig.~\ref{fig::R}, with deviations 
well within the statistical and systematical error of 3 to 5\%. 
Last not least, the result described above is in perfect agreement with
the recent lattice determination mentioned above.

Let us now discuss beauty, with $m_b$ evaluated at 
$\mu=10\;{\rm GeV}$. Again we first study the stability of the perturbative
expansion, subsequently the consistency of the experimental input.
With $\alpha_s(10\;{\rm GeV})=0.180$ the higher order corrections decrease
even more rapidly. 
Varying the scale $\mu$ between 5 and 15~GeV leads to a shift between 2 and 
6~MeV (Tab.~\ref{tab::mb}) which is completely negligible. 
Alternatively we may consider the analogue of eq.~(\ref{mc}) with the correction 
factor
\begin{equation}
\frac{m_b}{m_b^{\rm Born}}=
1 - 
\left(\begin{array}{c}0.270\\0.456\\0.546\\0.603\end{array}\right)\alpha_s - 
\left(\begin{array}{c}0.206\\0.272\\0.348\\0.410\end{array}\right)\alpha_s^2 +
\left(\begin{array}{c}-0.064\\0.048\\0.051\\0.012\end{array}\right)\alpha_s^3.
\end{equation}
Taking $r_n^{\rm max}(\alpha_s(10~\mbox{GeV}))^4$ for an error 
estimate leads to a relative error of 0.28 / 0.48 / 0.57 / 0.63 per mille 
for $n=1$, 2, 3 and 4, respectively, which is again smaller
than our previous estimate. 
Let us now move to a
critical discussion of the experimental input. The contribution from the
lowest four $\Upsilon$ resonances has been taken directly from 
PDG~\cite{Yao:2006px} with systematic errors of the lowest three added 
linearly. 
The analysis~\cite{Chetyrkin:2009fv} of a recent 
measurement~\cite{:2008hx} of $R_b$ in the threshold region 
up to 11.20~GeV has provided results consistent with the earlier
analysis \cite{Kuhn:2007vp} but has lead to a significant reduction of 
the error in $m_b$.

In comparison with the charm analysis a larger
contribution arises from the region where data are substituted by the 
theoretically predicted $R_b$ with relative contributions of 
63, 41, 26  and 17 percent for $n=1$, 2, 3 and 4 respectively.  
This is particularly valid for the lowest moment.
For this reason we prefer to use the result from
$n=2$, alternatively we could have also taken the one from $n=3$. Let us
now collect the arguments in favour of this approach:

{\it i)} 
For light and charmed quarks the prediction for $R$ based on 
pQCD works extremely well already two to three GeV above threshold. 
No systematic shift has been observed between theory and experiment,
in the case of massless quarks, starting from around 2~GeV, and for the 
cross section including charm at and above 5~GeV up to the bottom 
threshold (Fig.~\ref{fig::R}). 
It is thus highly unplausible that the same approach should fail for 
bottom production.

{\it ii)} 
If the true $R_b$ in the continuum above 11.2~GeV would
differ from the theory prediction by a sizable amount,  the
results for $n=1$, 2 and 3 would be mutually inconsistent. Specifically,
a shift of the continuum term by 5\% would move $m_b$, derived from
$n=1$,  2 and 3 by about 64~MeV,  21~MeV and  9~MeV respectively.

{\it To summarize:}
Charm and bottom quark mass determinations have made significant progress
during the past years. A further reduction of the theoretical and experimental
error seems difficult at present. However, independent experimental results on
the $R$-ratio would help to further consolidate the present situation.
The confirmation 
by a recent  lattice analysis with similarly small uncertainty gives 
additional confidence in the result for $m_c$.

\section*{Acknowledgments} 
This work was supported by the Deutsche Forschungsgemeinschaft through
the SFB/TR-9 ``Computational Particle Physics''.
The work of C.S. was partially supported by the
European Community's Marie Curie Research
Training Network {\it{Tools and Precision Calculations
for Physics Discoveries at Colliders}} (HEPTOOLS)
under contract MRNT-CT-2006-035505.
J.H.K. would like to thank the organizers of Quarks~2010 for their
hospitality.


\begin{thebibliography}{99}


%
%

\bibitem{Shifman:1978bx}
  M.~A.~Shifman, A.~I.~Vainshtein and V.~I.~Zakharov,
  Nucl.\ Phys.\  B {\bf 147} (1979) 385.

\bibitem{Kuhn:2001dm}
  J.~H.~K\"uhn and M.~Steinhauser,
  Nucl.\ Phys.\ B {\bf 619} (2001) 588
  [Erratum-ibid.\ B {\bf 640} (2002) 415]
  [arXiv:hep-ph/0109084].

\bibitem{Chetyrkin:1995ii}
  K.~G.~Chetyrkin, J.~H.~K\"uhn and M.~Steinhauser,
  Phys.\ Lett.\ B {\bf 371} (1996) 93
  [arXiv:hep-ph/9511430].

\bibitem{Chetyrkin:1996cf}
  K.~G.~Chetyrkin, J.~H.~K\"uhn and M.~Steinhauser,
  Nucl.\ Phys.\  B {\bf 482} (1996) 213
  [arXiv:hep-ph/9606230].

\bibitem{Chetyrkin:1997mb}
  K.~G.~Chetyrkin, J.~H.~K\"uhn and M.~Steinhauser,
  Nucl.\ Phys.\ B {\bf 505} (1997) 40
  [arXiv:hep-ph/9705254].

\bibitem{Kuhn:2007vp}
  J.~H.~K\"uhn, M.~Steinhauser and C.~Sturm,
  Nucl.\ Phys.\  B {\bf 778} (2007) 192
  [arXiv:hep-ph/0702103].

\bibitem{Chetyrkin:2006xg}
  K.~G.~Chetyrkin, J.~H.~K\"uhn and C.~Sturm,
  Eur.\ Phys.\ J.\ C {\bf 48} (2006) 107
  [arXiv:hep-ph/0604234].

\bibitem{Boughezal:2006px}
  R.~Boughezal, M.~Czakon and T.~Schutzmeier,
  Phys.\ Rev.\ D {\bf 74} (2006) 074006
  [arXiv:hep-ph/0605023].

\bibitem{Maier:2008he}
  A.~Maier, P.~Maierhofer and P.~Marqaurd,
  Phys.\ Lett.\  B {\bf 669} (2008) 88
  [arXiv:0806.3405 [hep-ph]].

\bibitem{Maier:2009fz}
  A.~Maier, P.~Maierhofer, P.~Marquard and A.~V.~Smirnov,
  Nucl.\ Phys.\  B {\bf 824} (2010) 1
  [arXiv:0907.2117 [hep-ph]].

\bibitem{Kiyo:2009gb}
  Y.~Kiyo, A.~Maier, P.~Maierhofer and P.~Marquard,
  Nucl.\ Phys.\  B {\bf 823} (2009) 269
  [arXiv:0907.2120 [hep-ph]].

\bibitem{Hoang:2008qy}
  A.~H.~Hoang, V.~Mateu and S.~Mohammad Zebarjad,
  Nucl.\ Phys.\  B {\bf 813} (2009) 349
  [arXiv:0807.4173 [hep-ph]].

\bibitem{Chetyrkin:2009fv}
  K.~G.~Chetyrkin, J.~H.~K\"uhn, A.~Maier, P.~Maierhofer, P.~Marquard,
  M.~Steinhauser and C.~Sturm,
  Phys.\ Rev.\  D {\bf 80} (2009) 074010
  [arXiv:0907.2110 [hep-ph]].

\bibitem{Boughezal:2006uu}
  R.~Boughezal, M.~Czakon and T.~Schutzmeier,
  Nucl.\ Phys.\ Proc.\ Suppl.\  {\bf 160} (2006) 160
  [arXiv:hep-ph/0607141].

\bibitem{Maier:2007yn}
  A.~Maier, P.~Maierhofer and P.~Marquard,
  Nucl.\ Phys.\  B {\bf 797} (2008) 218
  [arXiv:0711.2636 [hep-ph]].

\bibitem{Sturm:2008eb}
  C.~Sturm,
  JHEP {\bf 0809} (2008) 075
  [arXiv:0805.3358 [hep-ph]].

\bibitem{Allison:2008xk}
  I.~Allison {\it et al.}  [HPQCD Collaboration],
  Phys.\ Rev.\  D {\bf 78} (2008) 054513
  [arXiv:0805.2999 [hep-lat]].

\bibitem{Novikov:1977dq}
  V.~A.~Novikov, L.~B.~Okun, M.~A.~Shifman, A.~I.~Vainshtein, M.~B.~Voloshin and V.~I.~Zakharov,
  Phys.\ Rept.\  {\bf 41} (1978) 1.

\bibitem{Broadhurst:1994qj}
  D.~J.~Broadhurst, P.~A.~Baikov, V.~A.~Ilyin, J.~Fleischer, O.~V.~Tarasov and V.~A.~Smirnov,
%
  Phys.\ Lett.\ B {\bf 329} (1994) 103
  [arXiv:hep-ph/9403274].

\bibitem{Ioffe:2005ym}
  B.~L.~Ioffe,
  Prog.\ Part.\ Nucl.\ Phys.\  {\bf 56} (2006) 232
  [arXiv:hep-ph/0502148].

\bibitem{Chetyrkin:2000yt}
  K.~G.~Chetyrkin, J.~H.~K\"uhn and M.~Steinhauser,
  Comput.\ Phys.\ Commun.\  {\bf 133} (2000) 43
  [arXiv:hep-ph/0004189].

\bibitem{McNeile:2010ji}
  C.~McNeile, C.~T.~H.~Davies, E.~Follana, K.~Hornbostel and G.~P.~Lepage,
  Phys.\ Rev.\  D {\bf 82} (2010) 034512
  [arXiv:1004.4285 [hep-lat]].

\bibitem{Bodenstein:2010qx}
  S.~Bodenstein, J.~Bordes, C.~A.~Dominguez, J.~Penarrocha and K.~Schilcher,
  arXiv:1009.4325 [hep-ph].

\bibitem{Signer:2008da}
  A.~Signer,
  Phys.\ Lett.\  B {\bf 672} (2009) 333
  [arXiv:0810.1152 [hep-ph]].

\bibitem{Besson:1984bd}
  D.~Besson {\it et al.}  [CLEO Collaboration],
  Phys.\ Rev.\ Lett.\  {\bf 54} (1985) 381.

\bibitem{:2008hx}
  B.~Aubert {\it et al.}  [BABAR Collaboration],
  Phys.\ Rev.\ Lett.\  {\bf 102} (2009) 012001
  [arXiv:0809.4120 [hep-ex]].

\bibitem{Harlander:2002ur}
  R.~V.~Harlander and M.~Steinhauser,
  Comput.\ Phys.\ Commun.\  {\bf 153} (2003) 244
  [arXiv:hep-ph/0212294].

\bibitem{Yao:2006px}
  W.~M.~Yao {\it et al.}  [Particle Data Group],
  J.\ Phys.\ G {\bf 33} (2006) 1.



\end{thebibliography}
\end{document}